\documentclass[twocolumn,showpacs,preprintnumbers,amsmath,amssymb,superscriptaddress,prb]{revtex4}
\usepackage{graphicx}
\usepackage{longtable}
\usepackage{verbatim}

\usepackage{color}

\newcommand{\notekj}[1]{}
\begin{document}
\title{Band offsets at the crystalline/amorphous silicon interface from first-principles}
\author{K. Jarolimek}
\thanks{Present address: University of Kentucky, Center for Applied Energy Research,
2540 Research Park Drive, Lexington, KY 40511, USA}
\affiliation{Radboud University, Institute for Molecules and Materials, 
Heyendaalseweg 135, 6525 AJ Nijmegen, The Netherlands}

\author{E. Hazrati}
\affiliation{Radboud University, Institute for Molecules and Materials, 
Heyendaalseweg 135, 6525 AJ Nijmegen, The Netherlands}

\author{R. A. de Groot}
\affiliation{Radboud University, Institute for Molecules and Materials, 
Heyendaalseweg 135, 6525 AJ Nijmegen, The Netherlands}
\affiliation{University of Groningen, Zernike Institute for Advanced Materials,
Nijenborgh~4, 9747~AG~Groningen, The~Netherlands}

\author{G. A. de Wijs}
\affiliation{Radboud University, Institute for Molecules and Materials, 
Heyendaalseweg 135, 6525 AJ Nijmegen, The Netherlands}

\date{\today}

%73.21.Fg 	Quantum wells
%71.23.Cq 	Amorphous semiconductors, metallic glasses, glasses 
%71.15.Mb 	Density functional theory, local density approximation, gradient and other corrections
%71.15.Pd 	Molecular dynamics calculations (Car-Parrinello) and other numerical simulations 

\begin{abstract} 
The band offsets between crystalline and hydrogenated amorphous silicon (a-Si:H)
are key parameters governing the charge transport in modern silicon
hetrojunction solar cells. They are an important input for macroscopic
simulators that are used to further optimize the solar cell.
Past experimental studies, using X-ray photoelectron spectroscopy (XPS) and
capacitance-voltage measurements, have yielded conflicting results on the band
offset.
Here we present a computational study on the band offsets. It is based on
atomistic models and density-functional theory (DFT). The amorphous part of
the interface is obtained by relatively long DFT first-principles
molecular-dynamics (MD) runs at an elevated temperature on 30 statistically
independent samples.
In order to obtain a realistic conduction band position the
electronic structure of the interface is calculated
with a hybrid functional.
We find a
slight asymmetry in the band offsets, where the offset in the valence band
(0.30~eV) is larger than in the conduction band (0.17~eV). Our results are in
agreement with the latest XPS measurements that report a valence band offset of
0.3~eV [M. Liebhaber {\it et al.}, Appl. Phys. Lett. \textbf{106}, 031601 (2015)].
\end{abstract}
\maketitle

\section{Introduction} 
\label{sec_intro}
Silicon heterojunction (SHJ) solar cells combine the high-efficiency of c-Si wafer
technology with the high-throughput and low-cost of hydrogenated amorphous silicon
(a-Si:H) solar cells. The interface between crystalline and amorphous silicon
lies at the heart of the SHJ solar cell. Since a-Si:H has a larger band gap
than c-Si, band offsets are formed at the interface.

Experimentally the band offsets can be determined with techniques such as
photoelectron spectroscopy and capacitance-voltage measurements. The reported
values, however, scatter in a broad range.\cite{kleider2012}
This can be due to different
deposition conditions of the a-Si:H layer or misinterpretation of the
experimental results. On average it appears that the offset at the valence band
is larger than at the conduction band (Ref.~\onlinecite{kleider2012}, p.~418).

Theoretical studies were mostly concerned with the atomic structure of the
interface between c-Si and \emph{pure} a-Si. Studies aimed either to
obtain the interface energy (Ref.~\onlinecite{izumi2004}) or to study
the velocity of the crystallization of a-Si on
c-Si substrates.\cite{krzeminski2007} Some studies reported the electronic
density of states, for the c-Si and a-Si:H parts of the interface, but did
not comment on the band offsets.\cite{nolan2012,pan2004,tosolini2004}
Santos {\it et al.}\cite{santow2014} investigated defects present at the
interface and the corresponding electronic levels within the band gap.
George {\it et al.}\cite{george2013} modeled the electron spin resonance
signal of defects at the interface. In terms of band offset calculations
we are aware of two studies that obtained the values from the
respective bulk materials.\cite{allan1998,vandewalle1995}
Peressi {\it et al.},\cite{peressi2001} used a complete interface model,
prepared with a combination of
classical and first-principle molecular dynamics.\cite{peressi2001}
Their amorphous part, was however, build from pure a-Si and often contained
a high defect concentration making the interface semi-metallic.

We present a calculation of band offsets that improves upon published studies
in several aspects. The band offsets are calculated from an explicit interface
model and not extracted from bulk properties only. There is a substantial (30)
number of independent structural models that allows for reliable statistics.
These models are
prepared entirely from first-principles MD with defect levels
sufficiently low to determine the band edges. The electronic
structure is described with hybrid functionals that give a better description
of band gaps and conduction band offsets.

This paper is organized as follows. Section~\ref{sec_technical} provides
technical details on the calculations. The preparation of the structural models
is described in Sec.~\ref{sec_prep}. In Sec.~\ref{sec_elstruct} we discuss the
electronic structure of the interface. Conclusions are presented in
Sec.~\ref{sec_conclusions}.

\section{Technical details}
\label{sec_technical}
Calculations were performed on the level of density functional theory (DFT)
with the Vienna Ab initio Simulation Package
(VASP).\cite{kresse_1993,kresse_1996} Electron-ion interactions were described
using the projector augmented wave (PAW) method.\cite{blochl_1994, kresse_1999} 

We performed molecular dynamics calculations with the Verlet algorithm. The
canonical $NVT$ ensemble was simulated using the algorithm by
Nos\'{e}.\cite{nose1984} We increased the mass of hydrogen to 10~amu, which
allowed us to use a slightly longer time step of 1.5~fs. During the whole MD
run and the relaxation, we used the $\Gamma$ point for Brillouin zone sampling.
The kinetic energy cut-off was set relatively low at 150~eV. This was made
possible by using PAW potentials with larger core radii.\cite{hintzsche2012}
For Si the $s$-, $p$- and $d$-partial wave radii were 2.2, 2.7 and 2.7~a.u.,
respectively. For H both $s$- and $p$-partial wave radii were 1.3~a.u. The
performance of the potentials was tested on bulk c-Si and the SiH$_4$ molecule.
The equilibrium Si-Si and Si-H bond lengths decreased by less than 0.01~\AA,
when using the less accurate potentials. The frequency of the TO mode in c-Si
decreased by 1~\%, while the frequency of the Si-H stretching mode in SiH$_4$
was lower by 5~\%.  The above described tests, as well as all dynamic
calculations were performed with the generalized gradient approximation (GGA)
using the PBEsol functional.\cite{perdew2008}

Although the GGA gives accurate structural properties it is known to
underestimate band gaps. In order to obtain realistic band gaps and
offsets,\cite{heyd2005} all static calculations were performed with a hybrid
functional. This type of functional include a part of exact exchange from
Hartree-Fock theory.  We used the HSE06 hybrid functional\cite{Heyd2003,
Heyd2006} with a screening parameter of 0.2~\AA$^{-1}$. PBE potentials with a
250~eV cut-off were used. The Brillouin zone was sampled with a $\Gamma$
centered 2$\times$2$\times$1 mesh, while the Hartree-Fock kernel was evaluated
only at $\Gamma$. The density of states was calculated with a Gaussian
smearing with a width of 0.05~eV.

\section{Preparation of the structure}
\label{sec_prep}
\begin{figure}[tbp]
\centering
\includegraphics[angle=0, width=0.48\textwidth]{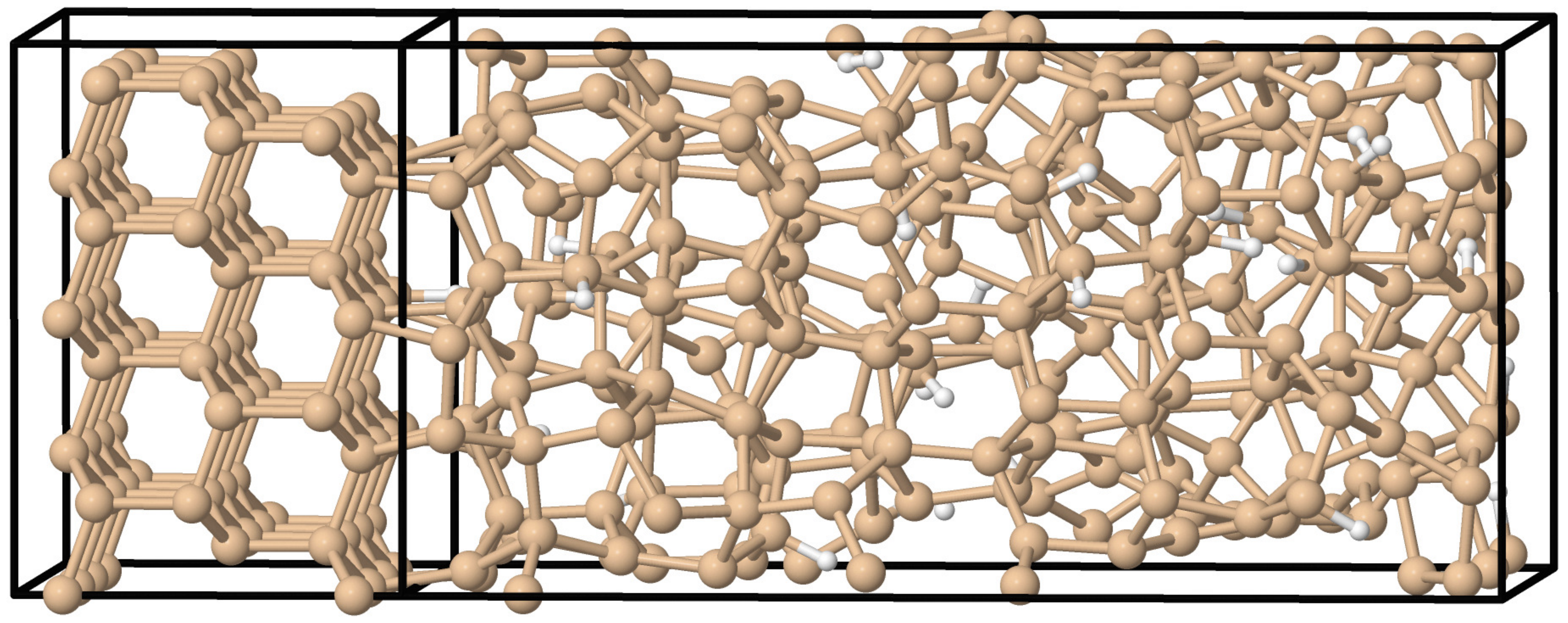}
\caption{(Color online) One of the simulation cells used in the present study.
The crystalline cell was terminated with two distinct (111) surfaces. Si and H
atoms are drawn as brown and white spheres, respectively.}
\label{fig_cell2}
\end{figure}

In order to simulate the interface we constructed a total of 30~simulation cells
with dimensions of 15.35$\times$13.30$\times$36.00~\AA$^3$. Periodic boundary
conditions were applied in all three dimensions. The cells were divided into a
crystalline and an amorphous part. The crystalline part consisted of 3~double
layers of Si atoms, centered within a 9.40~\AA\ wide region (see
Fig.~\ref{fig_cell2}). 

The amorphous part contains 256 Si atoms and 30 H atoms, which leads to a H
concentration of 10.5~at.~\% and mass density of 2.21~g/cm$^3$. These values
are representative of device-quality bulk a-Si:H.\cite{remes1998} Initially the
atoms were placed randomly in the cell, followed by an annealing step at
1100~K for 135~ps, using DFT molecular-dynamics. The optimum annealing
temperature was determined by a series of tests on bulk a-Si:H cells. Low
temperatures did not result in sufficient movement of atoms and the system could
be trapped in a high-energy local minimum. On the other hand, a too high
temperature could result in a liquid-like structure with many over-coordinated
defects.

After the annealing step a relaxation was performed that brought the system to
the nearest local energy minimum. During the annealing and relaxation, atoms in
the crystalline part were fixed and were not allowed to move. As a last step we
doubled the amount of c-Si in the simulation cell. The cell vector perpendicular
to the interface increased from 36.00 to 45.40~\AA.

In the following we analyze the structure of the middle portion of the
amorphous cell. Only atoms that were located more than 3.1~\AA\ from the nearest
interface were considered. The mean Si-Si and Si-H bond lengths were 2.37 and
1.53~\AA, respectively. These values compare well with diffraction measurements
on bulk a-Si:H, that give 2.35 and 1.48~\AA,
respectively.\cite{bellissent1989} On average we found 2.1 H-H bonds (H$_2$
molecules) per simulation cell. A more sensitive measure of strain in the
amorphous network is the bond angle distribution. To proceed, we define the
following cut-off distances: $r_\text{Si-Si} = 2.76$~\AA\ and $r_\text{Si-H} =
1.79$~\AA. The bond angles were calculated only between Si-Si bonds and Si-H
bonds were ignored. We obtained an average value of 108.9$^\circ$, that is close
to the bulk experimental value of 109.5$^\circ$.\cite{bellissent1989} The bond
angle RMS deviation 
can be inferred from the width of the TO peak, as measured by
Raman spectroscopy. The experimental values of 8.7$^\circ$
(Ref.~\onlinecite{vignoli2005}) and 9.3$^\circ$ (Ref.~\onlinecite{wakagi1994})
are a bit smaller than the calculated value of 12.6$^\circ$. This points to some
additional strain in our models, although one has to keep in mind that the
experiments were done on bulk a-Si:H. It is reasonable to assume that the
amorphous network is more strained close to the interface with c-Si. 

Another important property of the a-Si:H model is the number of coordination
defects. We find that the number of 5-fold coordinated Si atoms is higher than
the number 3-fold coordinated ones (4.6 and 1.4 atoms per simulation cell,
respectively). Tosolini {\it et~al.}\cite{tosolini2004} reported an opposite
trend and Santos {\it et~al.}\cite{santos2014} found 3-fold coordinated atoms
but did not report on the 5-fold coordinated ones. We also find H atoms in
a bridging position (0.2~atoms per simulation cell).
Santos {\it et~al.}\cite{santos2014} also investigated
this defect but it was not reported by Tosolini {\it et~al}.\cite{tosolini2004}
The number of other defects is less than 0.1 per simulation cell. 
The differences with Santos {\it et~al.} and Tosolini {\it et~al.} are possibly
due to the different method used to prepare the structure: they used
tight-binding MD whereas we used DFT MD.
The presence of the interface might lead to more 5-fold coordinated Si atoms.
In our previous DFT MD study on pure a-Si:H, which had similar
H concentration, we obtained comparable numbers of 3-fold and 5-fold coordinated
Si atoms.\cite{Jarolimek2009a}

\section{Electronic structure of the interface}
\label{sec_elstruct}
Because we had to average over 30 simulation cells we needed to align the
single-particle energy levels in the different cells. We used the mean potential
at core of the Si atoms in the mid-section of the crystalline part (2~layers,
32~atoms) for this purpose. The potential at a particular atomic site was
calculated with a unit test charge that has a radius of 0.989~\AA. 

\begin{figure}[tbp]
\centering
\includegraphics[angle=0, width=0.48\textwidth]{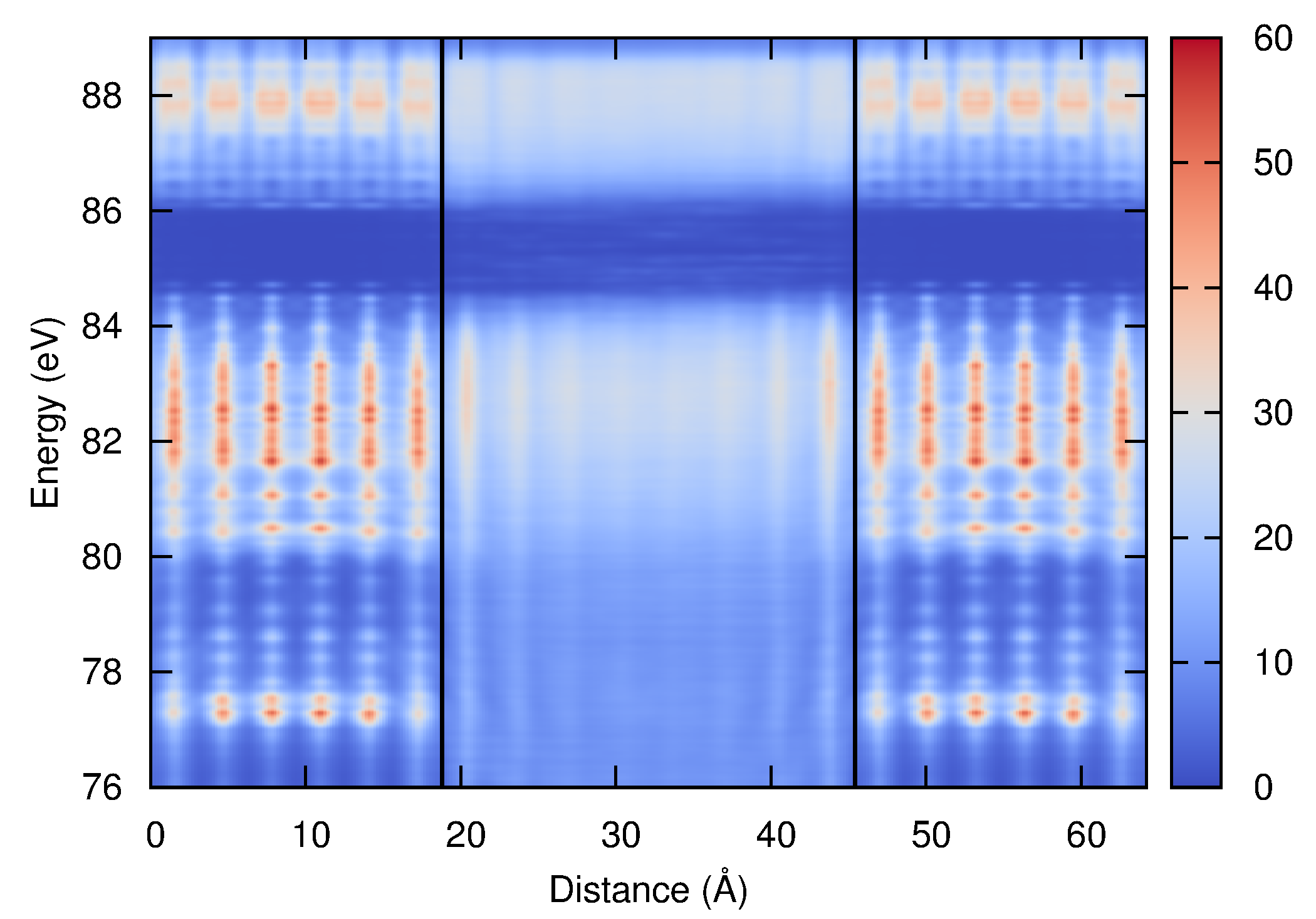}
\caption{(Color online) Position-resolved DOS (in 10$^{21}$~cm$^{-3}$eV$^{-1}$,
HSE06) along the normal of the interface and obtained as an average over
30~cells. The crystalline part of the cell is shown twice in order to see both
interfaces more clearly. Interfaces are marked by vertical lines. The zero of
energy is the mean potential at the Si atom cores in the central part of the
crystalline region.}
\label{fig_dosloc}
\end{figure}

In Fig.~\ref{fig_dosloc} we show position-resolved density of states of the interface.
There is a marked difference between the crystalline and amorphous part of the
interface. In the crystalline part the 6~double layers of Si atoms can be easily
identified as areas with high density of states. The disorder in the amorphous
part leads to a smeared out DOS. Some ordering is visible close to the
crystalline part. Evidently the ``memory'' of the layered structure is lost
only gradually when moving away from the fixed crystalline part.
The band gap is centered at around 85~eV and has a dark blue color.

\begin{figure}[tbp]
\centering
\includegraphics[angle=0, width=0.48\textwidth]{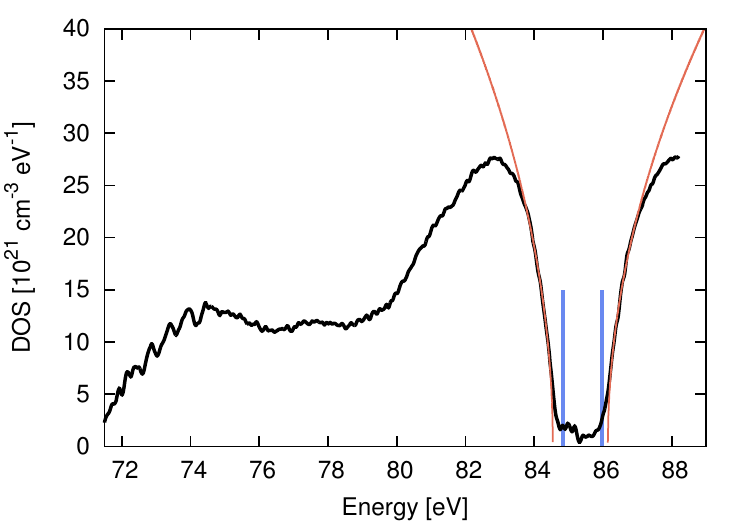}
\caption{(Color online) HSE06 DOS of the amorphous part of the interface (black
lines). Tauc fit to the DOS is shown in red. Position of band edges of c-Si are
marked with blue vertical lines.
The zero of energy is the mean potential at the Si atom cores in the central
part of the crystalline region.}
\label{fig_dos}
\end{figure}

In the following we consider the DOS of the middle section of the amorphous
part that is representative of bulk a-Si:H (see Fig.~\ref{fig_dos}). We discard
3.1~\AA\ (width of one double layer) from both edges of the amorphous part. 
The determination of a band gap for amorphous semiconductors is somewhat
ambiguous.\cite{O'Leary2004} We used the definition by
Tauc, which is the simplest one. In this model the valence and conduction band
DOS follow a square root dependence on energy (see red lines in
Fig.~\ref{fig_dos}). It is clear that the model is valid only for the middle
range of DOS values. We chose an interval that spans from 30 to 80~\% of the
maximum DOS value (at 28$\times$10$^{21}$~cm$^{-3}$eV$^{-1}$). This allowed us
to find the energy ranges to fit the Tauc model to the calculated DOS. Using a
lower limit of 30~\% effectively means that we rely on the extended states
to define the position of the band edge and that we suppress the effect of tail
and defect states. This should also minimize the effect of stress in the
structural models. After performing a least square fit we obtained a band gap of
1.60~eV (HSE06), which is quite close to the experimental Tauc gap of
1.7~eV.\cite{street1991a} We tested the sensitivity of the obtained band
edges on the number of cells used in the averaging. When using the first 10 or
20 cells the position of the band edges changed by less than 0.01~eV. Band
offsets were affected by the same amount.

\begin{figure}[tbp]
\centering
\includegraphics[angle=0, width=0.48\textwidth]{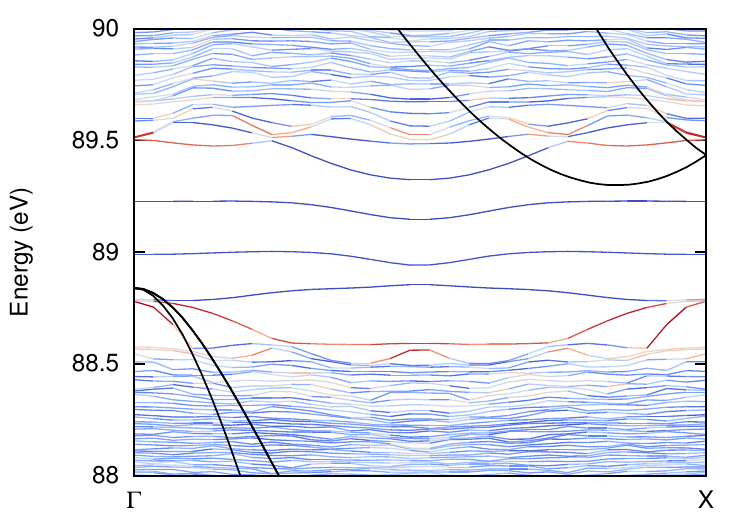}
\caption{(Color online) PBEsol Band structure of the c-Si/a-Si:H interface
(blue/red lines) compared to bulk c-Si (black lines). States marked with red
color (light grey) are localized in the crystalline part of the interface.
The zero of energy is the mean potential at the Si atom cores in the central
part of the crystalline region.}
\label{fig_bands}
\end{figure}

As a next step we needed to obtain the position of the c-Si band edges. One
might assume that one can do this by calculating the DOS of the crystalline
part of the interface, in an analogous way as was done for the amorphous part.
It turned out, however, that the band gap obtained in this way is overestimated.
We argue that this is due to quantum confinement effects. To illustrate this we
selected one of the cells and calculated its band structure with the GGA
functional (see Fig.~\ref{fig_bands}).
We chose a path in reciprocal space that is the same as for a primitive c-Si
cell with 2~atoms.\footnote{The primitive cell was cut from the interface cell
and thus the $\langle$111$\rangle$ vector points along the interface normal and
along the $z$-axis. This allows us to use the same Cartesian coordinates
(0.1303, 0.0752, 0.1064)~\AA$^{-1}$ for the X point, in both band structure
calculations.} The band structure of the interface cell is color coded so that
we can identify states (in red) that are localized in the crystalline part of
the interface.  We observe that the valence band maximum is still at $\Gamma$
and the conduction band minimum is close to the X point. In both cases the
bands are less dispersive than in bulk c-Si. The (PBEsol) band gap is larger
(0.69~eV) compared to bulk c-Si (0.46~eV). When we removed 3 double layers of
Si atoms
(9.40~\AA) from the crystalline part, we obtained an even larger band gap of
0.96~eV. Conversely adding 3 double layers decreased the band gap to 0.58~eV.
We note that we expect only weak quantum confinement and charge localization
effects in the amorphous part, since the charge carriers in bulk a-Si:H are
localized to begin with.\cite{jarolimek2014}

In order to circumvent the problem with quantum confinement in the c-Si part we
followed a different approach. We calculated the band structure of c-Si with
2~atoms in the unit cell and obtained a band gap of 1.14~eV (HSE06).
The position of the
top of the valence band (at $\Gamma$) and bottom of conduction band (between
$\Gamma$ and X) were again referenced to the potential at core of the Si atoms.
These values were then combined with the DOS of a-Si:H (see blue lines in
Fig.~\ref{fig_dos}). The band offsets were calculated as differences between
the positions of the band edges in c-Si and a-Si:H. For the valence and conduction
offset we obtained 0.30 and 0.17~eV, respectively. 

\section{Discussion}
\label{sec_discussion}
Our calculated valence band offset (0.30~eV) is in good agreement with
the latest X-ray photoelectron spectroscopy (XPS) measurement
% of the valence band offset
of 0.3~eV.\cite{liebhaber2015} Kleider performed a careful analysis of
capacitance-voltage measurements (C-V) and reports a valence and conduction offset of
0.40 and 0.15~eV, respectively (Ref.~\onlinecite{kleider2012}, p.~405).
Both are close to our values (0.30 and 0.17~eV). Kleider notes that
some of the previous studies did not take into account the specific properties
of a-Si:H in their analysis. This could then explain the large spread in
the reported values. 

Previous theoretical studies used models of bulk c-Si and a-Si:H to calculate
band offsets. Our results for the valence offset fall in-between the reported
values. Allan~{\it et~al.}\ used a $\sim$4000 atom model of a-Si:H and in the
tight-binding approximation obtained a valence offset of 0.36~eV. The H
concentration was 8~at.~\%, close to our value, but the mass density was not
reported. It is likely that the final value is, however, larger since the band
edge states used are somewhat localized (band tails). This is corroborated by
the fact that the reported band gap of a-Si:H is too small
(1.36~eV).\cite{allan1998} Van de Walle used the so called ``model-solid''
theory to obtain a valence band offset of 0.2~eV.\cite{vandewalle1995}
Van de Walle also derived relations between the offsets and a-Si:H density and H
concentration. The a-Si:H model has a similar H concentration (11~at.~\%) as
our models, but the density is higher (set to c-Si value). When we substitute
our density, to make a more fair comparison, the valence offset reduces
to 0.07~eV. It is difficult to pinpoint the origin of the discrepancies between
our results and the older studies. The fact that the interface was not
considered directly might play a role, together with differences in a-Si:H
structure preparation and density. In our view the most likely explanation is
that the methods to locate the band edges in a-Si:H were different. When small
cells or a small number of cells is used, it is difficult to distinguish between
extended and tail states. 

In terms of SHJ solar cell performance there seems to be an optimum value for
the valence band offset. Several studies simulated the performance of n-type
wafer based SHJ cells by solving the Poisson and charge carrier continuity
equations (see Ref.~\onlinecite{shen2013} and references therein). The studies
agree that a valence offset larger than 0.5~eV leads to a sharp decrease in
solar cell efficiency. This is caused by accumulation and subsequent
recombination of holes on the c-Si side of the c-Si/a-Si:H interface.
Shen~{\it et~al.}\ put the optimum offset at 0.45~eV.\cite{shen2013}

\section{Conclusions}
\label{sec_conclusions}
We have prepared an atomistic model of the crystalline/amorphous silicon
interface. The amorphous part is hydrogenated and is thus relevant for
technological applications such as silicon heterojunction solar cells. In order
to obtain reliable results we averaged the calculated quantities over 30
statistically independent simulations cell. Atomic models were prepared with
molecular-dynamics, where forces are computed with density functional theory.
This should give an accurate description of the interface region that contains
a large number of strained bonds.  

The electronic structures of the particular simulation cells are aligned at the
mean potential at the Si atom cores in the crystalline part of the interface. We
have attempted to extract the band offsets directly from the position resolved
density of states. This was possible in the amorphous part but not in the
crystalline part of the interface due to quantum confinement effects. To
resolve this issue we used band edges from a bulk c-Si calculation and performed
again an alignment at the core potential of Si atoms. We obtain a valence and
conduction band offset of 0.30 and 0.17~eV. This is in good agreement with
recent XPS and C-V measurements.\cite{liebhaber2015,kleider2012}

\begin{acknowledgments}
We thank Prof.~Kresse for the special, extra-soft PAW data sets.
This work was supported by Technology Foundation STW within the framework of
the FLASH perspective program. We thank SURFsara (www.surfsara.nl) for the
support in using the Lisa Compute Cluster. The work of RAG is part of the
research program of the Foundation for Fundamental Research on Matter (FOM).
Both SURFsara and FOM are financially supported by the Netherlands Organization
for Scientific Research (NWO). 
\end{acknowledgments}

\bibliography{library}

\begin{thebibliography}{31}
\expandafter\ifx\csname natexlab\endcsname\relax\def\natexlab#1{#1}\fi
\expandafter\ifx\csname bibnamefont\endcsname\relax
  \def\bibnamefont#1{#1}\fi
\expandafter\ifx\csname bibfnamefont\endcsname\relax
  \def\bibfnamefont#1{#1}\fi
\expandafter\ifx\csname citenamefont\endcsname\relax
  \def\citenamefont#1{#1}\fi
\expandafter\ifx\csname url\endcsname\relax
  \def\url#1{\texttt{#1}}\fi
\expandafter\ifx\csname urlprefix\endcsname\relax\def\urlprefix{URL }\fi
\providecommand{\bibinfo}[2]{#2}
\providecommand{\eprint}[2][]{\url{#2}}

\bibitem[{\citenamefont{Kleider}(2012)}]{kleider2012}
\bibinfo{author}{\bibfnamefont{J.-P.} \bibnamefont{Kleider}}, in
  \emph{\bibinfo{booktitle}{Physics and Technology of Amorphous-Crystalline
  Heterostructure Silicon Solar Cells}}, edited by
  \bibinfo{editor}{\bibfnamefont{W.~G. J. H.~M.} \bibnamefont{van Sark}},
  \bibinfo{editor}{\bibfnamefont{L.}~\bibnamefont{Korte}}, \bibnamefont{and}
  \bibinfo{editor}{\bibfnamefont{F.}~\bibnamefont{Roca}}
  (\bibinfo{publisher}{Springer}, \bibinfo{address}{Berlin},
  \bibinfo{year}{2012}).

\bibitem[{\citenamefont{Izumi et~al.}(2004)\citenamefont{Izumi, Hara, Kumagai,
  and Sakai}}]{izumi2004}
\bibinfo{author}{\bibfnamefont{S.}~\bibnamefont{Izumi}},
  \bibinfo{author}{\bibfnamefont{S.}~\bibnamefont{Hara}},
  \bibinfo{author}{\bibfnamefont{T.}~\bibnamefont{Kumagai}}, \bibnamefont{and}
  \bibinfo{author}{\bibfnamefont{S.}~\bibnamefont{Sakai}},
  \bibinfo{journal}{Comp. Mater. Sci.} \textbf{\bibinfo{volume}{31}},
  \bibinfo{pages}{279} (\bibinfo{year}{2004}).

\bibitem[{\citenamefont{Krzeminski et~al.}(2007)\citenamefont{Krzeminski,
  Brulin, Cuny, Lecat, Lampin, and Cleri}}]{krzeminski2007}
\bibinfo{author}{\bibfnamefont{C.}~\bibnamefont{Krzeminski}},
  \bibinfo{author}{\bibfnamefont{Q.}~\bibnamefont{Brulin}},
  \bibinfo{author}{\bibfnamefont{V.}~\bibnamefont{Cuny}},
  \bibinfo{author}{\bibfnamefont{E.}~\bibnamefont{Lecat}},
  \bibinfo{author}{\bibfnamefont{E.}~\bibnamefont{Lampin}}, \bibnamefont{and}
  \bibinfo{author}{\bibfnamefont{F.}~\bibnamefont{Cleri}}, \bibinfo{journal}{J.
  Appl. Phys.} \textbf{\bibinfo{volume}{101}}, \bibinfo{pages}{123506}
  (\bibinfo{year}{2007}).

\bibitem[{\citenamefont{Nolan et~al.}(2012)\citenamefont{Nolan, Legesse, and
  Fagas}}]{nolan2012}
\bibinfo{author}{\bibfnamefont{M.}~\bibnamefont{Nolan}},
  \bibinfo{author}{\bibfnamefont{M.}~\bibnamefont{Legesse}}, \bibnamefont{and}
  \bibinfo{author}{\bibfnamefont{G.}~\bibnamefont{Fagas}},
  \bibinfo{journal}{Phys. Chem. Chem. Phys.} \textbf{\bibinfo{volume}{14}},
  \bibinfo{pages}{15173} (\bibinfo{year}{2012}).

\bibitem[{\citenamefont{Pan and Biswas}(2004)}]{pan2004}
\bibinfo{author}{\bibfnamefont{B.~C.} \bibnamefont{Pan}} \bibnamefont{and}
  \bibinfo{author}{\bibfnamefont{R.}~\bibnamefont{Biswas}},
  \bibinfo{journal}{J. Appl. Phys.} \textbf{\bibinfo{volume}{96}},
  \bibinfo{pages}{6247} (\bibinfo{year}{2004}).

\bibitem[{\citenamefont{Tosolini et~al.}(2004)\citenamefont{Tosolini, Colombo,
  and Peressi}}]{tosolini2004}
\bibinfo{author}{\bibfnamefont{M.}~\bibnamefont{Tosolini}},
  \bibinfo{author}{\bibfnamefont{L.}~\bibnamefont{Colombo}}, \bibnamefont{and}
  \bibinfo{author}{\bibfnamefont{M.}~\bibnamefont{Peressi}},
  \bibinfo{journal}{Phys. Rev. B} \textbf{\bibinfo{volume}{69}},
  \bibinfo{pages}{075301} (\bibinfo{year}{2004}).

\bibitem[{\citenamefont{George et~al.}(2013)\citenamefont{George, Behrends,
  Schnegg, Schulze, Fehr, Korte, Rech, Lips, Rohrm\"{u}ller, Rauls
  et~al.}}]{george2013}
\bibinfo{author}{\bibfnamefont{B.~M.} \bibnamefont{George}},
  \bibinfo{author}{\bibfnamefont{J.}~\bibnamefont{Behrends}},
  \bibinfo{author}{\bibfnamefont{A.}~\bibnamefont{Schnegg}},
  \bibinfo{author}{\bibfnamefont{T.~F.} \bibnamefont{Schulze}},
  \bibinfo{author}{\bibfnamefont{M.}~\bibnamefont{Fehr}},
  \bibinfo{author}{\bibfnamefont{L.}~\bibnamefont{Korte}},
  \bibinfo{author}{\bibfnamefont{B.}~\bibnamefont{Rech}},
  \bibinfo{author}{\bibfnamefont{K.}~\bibnamefont{Lips}},
  \bibinfo{author}{\bibfnamefont{M.}~\bibnamefont{Rohrm\"{u}ller}},
  \bibinfo{author}{\bibfnamefont{E.}~\bibnamefont{Rauls}},
  \bibnamefont{et~al.}, \bibinfo{journal}{Phys. Rev. Lett.}
  \textbf{\bibinfo{volume}{110}}, \bibinfo{pages}{136803}
  (\bibinfo{year}{2013}).

\bibitem[{\citenamefont{Allan et~al.}(1998)\citenamefont{Allan, Delerue, and
  Lannoo}}]{allan1998}
\bibinfo{author}{\bibfnamefont{G.}~\bibnamefont{Allan}},
  \bibinfo{author}{\bibfnamefont{C.}~\bibnamefont{Delerue}}, \bibnamefont{and}
  \bibinfo{author}{\bibfnamefont{M.}~\bibnamefont{Lannoo}},
  \bibinfo{journal}{Phys. Rev. B} \textbf{\bibinfo{volume}{57}},
  \bibinfo{pages}{6933} (\bibinfo{year}{1998}).

\bibitem[{\citenamefont{Van~de Walle and Yang}(1995)}]{vandewalle1995}
\bibinfo{author}{\bibfnamefont{C.~G.} \bibnamefont{Van~de Walle}}
  \bibnamefont{and} \bibinfo{author}{\bibfnamefont{L.~H.} \bibnamefont{Yang}},
  \bibinfo{journal}{J. Vac. Sci. Technol. B} \textbf{\bibinfo{volume}{13}},
  \bibinfo{pages}{1635} (\bibinfo{year}{1995}).

\bibitem[{\citenamefont{Peressi et~al.}(2001)\citenamefont{Peressi, Colombo,
  and de~Gironcoli}}]{peressi2001}
\bibinfo{author}{\bibfnamefont{M.}~\bibnamefont{Peressi}},
  \bibinfo{author}{\bibfnamefont{L.}~\bibnamefont{Colombo}}, \bibnamefont{and}
  \bibinfo{author}{\bibfnamefont{S.}~\bibnamefont{de~Gironcoli}},
  \bibinfo{journal}{Phys. Rev. B} \textbf{\bibinfo{volume}{64}},
  \bibinfo{pages}{193303} (\bibinfo{year}{2001}).

\bibitem[{\citenamefont{Kresse and Hafner}(1993)}]{kresse_1993}
\bibinfo{author}{\bibfnamefont{G.}~\bibnamefont{Kresse}} \bibnamefont{and}
  \bibinfo{author}{\bibfnamefont{J.}~\bibnamefont{Hafner}},
  \bibinfo{journal}{Phys. Rev. B} \textbf{\bibinfo{volume}{47}},
  \bibinfo{pages}{558} (\bibinfo{year}{1993}).

\bibitem[{\citenamefont{Kresse and Furthm\"{u}ller}(1996)}]{kresse_1996}
\bibinfo{author}{\bibfnamefont{G.}~\bibnamefont{Kresse}} \bibnamefont{and}
  \bibinfo{author}{\bibfnamefont{J.}~\bibnamefont{Furthm\"{u}ller}},
  \bibinfo{journal}{Phys. Rev. B} \textbf{\bibinfo{volume}{54}},
  \bibinfo{pages}{11169} (\bibinfo{year}{1996}).

\bibitem[{\citenamefont{Bl\"{o}chl}(1994)}]{blochl_1994}
\bibinfo{author}{\bibfnamefont{P.~E.} \bibnamefont{Bl\"{o}chl}},
  \bibinfo{journal}{Phys. Rev. B} \textbf{\bibinfo{volume}{50}},
  \bibinfo{pages}{17953} (\bibinfo{year}{1994}).

\bibitem[{\citenamefont{Kresse and Joubert}(1999)}]{kresse_1999}
\bibinfo{author}{\bibfnamefont{G.}~\bibnamefont{Kresse}} \bibnamefont{and}
  \bibinfo{author}{\bibfnamefont{D.}~\bibnamefont{Joubert}},
  \bibinfo{journal}{Phys. Rev. B} \textbf{\bibinfo{volume}{59}},
  \bibinfo{pages}{1758} (\bibinfo{year}{1999}).

\bibitem[{\citenamefont{Nos\'{e}}(1984)}]{nose1984}
\bibinfo{author}{\bibfnamefont{S.}~\bibnamefont{Nos\'{e}}},
  \bibinfo{journal}{J. Chem. Phys.} \textbf{\bibinfo{volume}{81}},
  \bibinfo{pages}{511} (\bibinfo{year}{1984}).

\bibitem[{\citenamefont{Hintzsche et~al.}(2012)\citenamefont{Hintzsche, Fang,
  Watts, Marsman, Jordan, Lamers, Weeber, and Kresse}}]{hintzsche2012}
\bibinfo{author}{\bibfnamefont{L.~E.} \bibnamefont{Hintzsche}},
  \bibinfo{author}{\bibfnamefont{C.~M.} \bibnamefont{Fang}},
  \bibinfo{author}{\bibfnamefont{T.}~\bibnamefont{Watts}},
  \bibinfo{author}{\bibfnamefont{M.}~\bibnamefont{Marsman}},
  \bibinfo{author}{\bibfnamefont{G.}~\bibnamefont{Jordan}},
  \bibinfo{author}{\bibfnamefont{M.~W. P.~E.} \bibnamefont{Lamers}},
  \bibinfo{author}{\bibfnamefont{A.~W.} \bibnamefont{Weeber}},
  \bibnamefont{and} \bibinfo{author}{\bibfnamefont{G.}~\bibnamefont{Kresse}},
  \bibinfo{journal}{Phys. Rev. B} \textbf{\bibinfo{volume}{86}},
  \bibinfo{pages}{235204} (\bibinfo{year}{2012}).

\bibitem[{\citenamefont{Perdew et~al.}(2008)\citenamefont{Perdew, Ruzsinszky,
  Csonka, Vydrov, Scuseria, Constantin, Zhou, and Burke}}]{perdew2008}
\bibinfo{author}{\bibfnamefont{J.~P.} \bibnamefont{Perdew}},
  \bibinfo{author}{\bibfnamefont{A.}~\bibnamefont{Ruzsinszky}},
  \bibinfo{author}{\bibfnamefont{G.~I.} \bibnamefont{Csonka}},
  \bibinfo{author}{\bibfnamefont{O.~A.} \bibnamefont{Vydrov}},
  \bibinfo{author}{\bibfnamefont{G.~E.} \bibnamefont{Scuseria}},
  \bibinfo{author}{\bibfnamefont{L.~A.} \bibnamefont{Constantin}},
  \bibinfo{author}{\bibfnamefont{X.}~\bibnamefont{Zhou}}, \bibnamefont{and}
  \bibinfo{author}{\bibfnamefont{K.}~\bibnamefont{Burke}},
  \bibinfo{journal}{Phys. Rev. Lett.} \textbf{\bibinfo{volume}{100}},
  \bibinfo{pages}{136406} (\bibinfo{year}{2008}).

\bibitem[{\citenamefont{Heyd et~al.}(2005)\citenamefont{Heyd, Peralta,
  Scuseria, and Martin}}]{heyd2005}
\bibinfo{author}{\bibfnamefont{J.}~\bibnamefont{Heyd}},
  \bibinfo{author}{\bibfnamefont{J.~E.} \bibnamefont{Peralta}},
  \bibinfo{author}{\bibfnamefont{G.~E.} \bibnamefont{Scuseria}},
  \bibnamefont{and} \bibinfo{author}{\bibfnamefont{R.~L.}
  \bibnamefont{Martin}}, \bibinfo{journal}{J. Chem. Phys.}
  \textbf{\bibinfo{volume}{123}}, \bibinfo{pages}{174101}
  (\bibinfo{year}{2005}).

\bibitem[{\citenamefont{Heyd et~al.}(2003)\citenamefont{Heyd, Scuseria, and
  Ernzerhof}}]{Heyd2003}
\bibinfo{author}{\bibfnamefont{J.}~\bibnamefont{Heyd}},
  \bibinfo{author}{\bibfnamefont{G.~E.} \bibnamefont{Scuseria}},
  \bibnamefont{and}
  \bibinfo{author}{\bibfnamefont{M.}~\bibnamefont{Ernzerhof}},
  \bibinfo{journal}{J. Chem. Phys.} \textbf{\bibinfo{volume}{118}},
  \bibinfo{pages}{8207} (\bibinfo{year}{2003}).

\bibitem[{\citenamefont{Heyd et~al.}(2006)\citenamefont{Heyd, Scuseria, and
  Ernzerhof}}]{Heyd2006}
\bibinfo{author}{\bibfnamefont{J.}~\bibnamefont{Heyd}},
  \bibinfo{author}{\bibfnamefont{G.~E.} \bibnamefont{Scuseria}},
  \bibnamefont{and}
  \bibinfo{author}{\bibfnamefont{M.}~\bibnamefont{Ernzerhof}},
  \bibinfo{journal}{J. Chem. Phys.} \textbf{\bibinfo{volume}{124}},
  \bibinfo{pages}{219906} (\bibinfo{year}{2006}).

\bibitem[{\citenamefont{Reme\v{s} et~al.}(1998)\citenamefont{Reme\v{s},
  Van\v{e}\v{c}ek, Torres, Kroll, Mahan, and Crandall}}]{remes1998}
\bibinfo{author}{\bibfnamefont{Z.}~\bibnamefont{Reme\v{s}}},
  \bibinfo{author}{\bibfnamefont{M.}~\bibnamefont{Van\v{e}\v{c}ek}},
  \bibinfo{author}{\bibfnamefont{P.}~\bibnamefont{Torres}},
  \bibinfo{author}{\bibfnamefont{U.}~\bibnamefont{Kroll}},
  \bibinfo{author}{\bibfnamefont{A.~H.} \bibnamefont{Mahan}}, \bibnamefont{and}
  \bibinfo{author}{\bibfnamefont{R.~S.} \bibnamefont{Crandall}},
  \bibinfo{journal}{J. Non-Cryst. Solids} \textbf{\bibinfo{volume}{227}},
  \bibinfo{pages}{876} (\bibinfo{year}{1998}).

\bibitem[{\citenamefont{Bellissent et~al.}(1989)\citenamefont{Bellissent,
  Menelle, Howells, Wright, Brunier, Sinclair, and Jansen}}]{bellissent1989}
\bibinfo{author}{\bibfnamefont{R.}~\bibnamefont{Bellissent}},
  \bibinfo{author}{\bibfnamefont{A.}~\bibnamefont{Menelle}},
  \bibinfo{author}{\bibfnamefont{W.~S.} \bibnamefont{Howells}},
  \bibinfo{author}{\bibfnamefont{A.~C.} \bibnamefont{Wright}},
  \bibinfo{author}{\bibfnamefont{T.~M.} \bibnamefont{Brunier}},
  \bibinfo{author}{\bibfnamefont{R.~N.} \bibnamefont{Sinclair}},
  \bibnamefont{and} \bibinfo{author}{\bibfnamefont{F.}~\bibnamefont{Jansen}},
  \bibinfo{journal}{Physica B} \textbf{\bibinfo{volume}{156}},
  \bibinfo{pages}{217} (\bibinfo{year}{1989}).

\bibitem[{\citenamefont{Vignoli et~al.}(2005)\citenamefont{Vignoli,
  M\'{e}linon, Masenelli, i~Cabarrocas, Flank, and Longeaud}}]{vignoli2005}
\bibinfo{author}{\bibfnamefont{S.}~\bibnamefont{Vignoli}},
  \bibinfo{author}{\bibfnamefont{P.}~\bibnamefont{M\'{e}linon}},
  \bibinfo{author}{\bibfnamefont{B.}~\bibnamefont{Masenelli}},
  \bibinfo{author}{\bibfnamefont{P.~R.} \bibnamefont{i~Cabarrocas}},
  \bibinfo{author}{\bibfnamefont{A.~M.} \bibnamefont{Flank}}, \bibnamefont{and}
  \bibinfo{author}{\bibfnamefont{C.}~\bibnamefont{Longeaud}},
  \bibinfo{journal}{J. Phys.: Condens. Matter} \textbf{\bibinfo{volume}{17}},
  \bibinfo{pages}{1279} (\bibinfo{year}{2005}).

\bibitem[{\citenamefont{Wakagi et~al.}(1994)\citenamefont{Wakagi, Ogata, and
  Nakano}}]{wakagi1994}
\bibinfo{author}{\bibfnamefont{M.}~\bibnamefont{Wakagi}},
  \bibinfo{author}{\bibfnamefont{K.}~\bibnamefont{Ogata}}, \bibnamefont{and}
  \bibinfo{author}{\bibfnamefont{A.}~\bibnamefont{Nakano}},
  \bibinfo{journal}{Phys. Rev. B} \textbf{\bibinfo{volume}{50}},
  \bibinfo{pages}{10666} (\bibinfo{year}{1994}).

\bibitem[{\citenamefont{Santos et~al.}(2014)\citenamefont{Santos, Cazzaniga,
  Onida, and Colombo}}]{santos2014}
\bibinfo{author}{\bibfnamefont{I.}~\bibnamefont{Santos}},
  \bibinfo{author}{\bibfnamefont{M.}~\bibnamefont{Cazzaniga}},
  \bibinfo{author}{\bibfnamefont{G.}~\bibnamefont{Onida}}, \bibnamefont{and}
  \bibinfo{author}{\bibfnamefont{L.}~\bibnamefont{Colombo}},
  \bibinfo{journal}{J. Phys.: Condens. Matter} \textbf{\bibinfo{volume}{26}},
  \bibinfo{pages}{095001} (\bibinfo{year}{2014}).

\bibitem[{\citenamefont{Jarolimek et~al.}(2009)\citenamefont{Jarolimek,
  de~Groot, de~Wijs, and Zeman}}]{Jarolimek2009a}
\bibinfo{author}{\bibfnamefont{K.}~\bibnamefont{Jarolimek}},
  \bibinfo{author}{\bibfnamefont{R.}~\bibnamefont{de~Groot}},
  \bibinfo{author}{\bibfnamefont{G.~A.} \bibnamefont{de~Wijs}},
  \bibnamefont{and} \bibinfo{author}{\bibfnamefont{M.}~\bibnamefont{Zeman}},
  \bibinfo{journal}{Phys. Rev. B} \textbf{\bibinfo{volume}{79}},
  \bibinfo{pages}{155206} (\bibinfo{year}{2009}).

\bibitem[{\citenamefont{O'Leary}(2004)}]{O'Leary2004}
\bibinfo{author}{\bibfnamefont{S.~K.} \bibnamefont{O'Leary}},
  \bibinfo{journal}{J. Mater. Sci.: Mater. Electron.}
  \textbf{\bibinfo{volume}{15}}, \bibinfo{pages}{401} (\bibinfo{year}{2004}).

\bibitem[{\citenamefont{Street}(1991)}]{street1991a}
\bibinfo{author}{\bibfnamefont{R.~A.} \bibnamefont{Street}},
  \emph{\bibinfo{title}{{Hydrogenated amorphous silicon}}}
  (\bibinfo{publisher}{Cambridge University Press},
  \bibinfo{address}{Cambridge}, \bibinfo{year}{1991}).

\bibitem[{\citenamefont{Jarolimek et~al.}(2014)\citenamefont{Jarolimek,
  de~Groot, de~Wijs, and Zeman}}]{jarolimek2014}
\bibinfo{author}{\bibfnamefont{K.}~\bibnamefont{Jarolimek}},
  \bibinfo{author}{\bibfnamefont{R.~A.} \bibnamefont{de~Groot}},
  \bibinfo{author}{\bibfnamefont{G.~A.} \bibnamefont{de~Wijs}},
  \bibnamefont{and} \bibinfo{author}{\bibfnamefont{M.}~\bibnamefont{Zeman}},
  \bibinfo{journal}{Phys. Rev. B} \textbf{\bibinfo{volume}{90}},
  \bibinfo{pages}{125430} (\bibinfo{year}{2014}).

\bibitem[{\citenamefont{Liebhaber et~al.}(2015)\citenamefont{Liebhaber, Mews,
  Schulze, Korte, Rech, and Lips}}]{liebhaber2015}
\bibinfo{author}{\bibfnamefont{M.}~\bibnamefont{Liebhaber}},
  \bibinfo{author}{\bibfnamefont{M.}~\bibnamefont{Mews}},
  \bibinfo{author}{\bibfnamefont{T.~F.} \bibnamefont{Schulze}},
  \bibinfo{author}{\bibfnamefont{L.}~\bibnamefont{Korte}},
  \bibinfo{author}{\bibfnamefont{B.}~\bibnamefont{Rech}}, \bibnamefont{and}
  \bibinfo{author}{\bibfnamefont{K.}~\bibnamefont{Lips}},
  \bibinfo{journal}{Appl. Phys. Lett.} \textbf{\bibinfo{volume}{106}},
  \bibinfo{pages}{031601} (\bibinfo{year}{2015}).

\bibitem[{\citenamefont{Shen et~al.}(2013)\citenamefont{Shen, Meng, and
  Liu}}]{shen2013}
\bibinfo{author}{\bibfnamefont{L.}~\bibnamefont{Shen}},
  \bibinfo{author}{\bibfnamefont{F.}~\bibnamefont{Meng}}, \bibnamefont{and}
  \bibinfo{author}{\bibfnamefont{Z.}~\bibnamefont{Liu}},
  \bibinfo{journal}{Solar Energy} \textbf{\bibinfo{volume}{97}},
  \bibinfo{pages}{168} (\bibinfo{year}{2013}).

\end{thebibliography}

\end{document}